\documentclass{PoS}

\usepackage{aas_macros}
\usepackage{floatrow}
\title{Changing-look Narrow-Line Seyfert 1s?}

\ShortTitle{ Changing-look NLS1s?}

\author{\speaker{Victor L. Oknyansky}$^1$, Konstantin L. Malanchev$^{1,2}$
and C. Martin Gaskell $^3$\\
\llap{$^1$}Sternberg Astronomical Institute, M.~V.~Lomonosov Moscow State University,
Moscow 119234, Russia\\
\llap{$^2$}National Research University Higher School of Economics,
Moscow 105066, Russia\\
\llap{$^3$}Department of Astronomy and Astrophysics, University of California, Santa Cruz, CA 95064, USA\\
\\
E-mail: \email{oknyan@mail.ru}, \email{malanchev@sai.msu.ru}, \email{mgaskell@ucsc.edu} }

\abstract{Two major challenges to unification schemes for active galactic nuclei (AGN) are the existence of Narrow-Line Seyfert 1s (NLS1s) and the existence of changing-look (CL) AGNs.
AGNs can drastically change their spectral appearance in the optical (changing their Seyfert type) and/or in the X-ray region. We illustrate the CL phenomenon with our multi-wavelength monitoring of  NGC 2617 and discuss its properties compared with NLS1s. There are few examples of CL NLS1s and the changes are mostly only in the X-ray region. 
It has been proposed that some of these could be cases of a tidal-disruption events (TDE) or supernova events. If BLRs have a flat geometry and NLS1s are seen face-on
then we have to see CL cases only if the orientation of the BLR changes as a result of a TDE or a close encounter of a star without a TDE.
If NLS1s include both high Eddington  accretion rate and low-inclination AGNs
then a significant fraction of NLS1s could be obscured and would not be identified as NLS1s. CL cases might happen more in such objects if dust sublimation occurs following a strong increase in the optical luminosity.}

\FullConference{Revisiting narrow-line Seyfert 1 galaxies and their place in the Universe - NLS1 Padova \\
		9-13 April 2018 \\
		Padova Botanical Garden, Italy}

\begin{document}

\bibliographystyle{jhep}

\section{Introduction}

Despite the successes of simple orientation-based AGN unification scheme, there are
significant problems that cannot be explained solely by different orientations. Major challenges to the simple model are:\\
1)The existence of the Narrow Line Seyfert 1s (NLS1s; \cite{Gaskell84,Osterbrock85}).\\
2) The existence of ``changing-look'' AGNs (CL AGNs).  A well-known typical example is NGC~4151 \cite{Lyuty84}.

In addition to the question of how to fit these objects into a unified scheme, there is an obvious question: how are NLS1s and CL AGNs related? This can tell us important things about both CL AGNs and about NLS1s.
  If BLR in NLS1 has a flattened geometry \cite{Gaskell09}  and they are seen close to face-on we would see a ``changing-look" only if the orientation of the BLR and accretion disc changes because of a major disruption such as a tidal-disruption event (TDE), a close encounter with the system of a stars without complete disruption, or a close passage of a secondary black hole. It has been argued that what we call NLS1s include both high Eddington ratio AGNs and low-inclination AGNs (e.g., \cite{Peterson11}).  In this case a significant fraction of NLS1s could be obscured and would mostly not be identified as NLS1s due to the difficulty of recognizing them from optical spectra.

As it was suspected by Dewangan et al.\cite{Dewangan05}, there must be Sy2s with high relative
accretion rates similar to those of NLS1s - that is, the type 2 counterpart of NLS1s. Some objects of this type have been recognized using near-infrared spectroscopy, optical spectropolarimetry, and X-ray observations: NGC 7314, 7582, 5506 and  Ark 1388 \cite{Doi15}. We could therefore expect that, as with known CL transitions between type-1 and type-2 states, some type-2 NLS1s might change to normal NLS1s if the obscuring dust clouds change. So far, however, the number of obscured (type-2) NLS1s known is very low and no such CL cases have been observed or recognized.

CL AGNs cases can be explained with either changing obscuration and a change in accretion rates. We have suggested \cite{Oknyansky17} that there can be a combination of these two factors: i.e., that a change in the energy-generation rate leads to a change in absorption of energy by dust. In our proposed hollow bi-conical dust distribution model \cite{Oknyansky15} (see Fig.1), sublimation or recreation of dust in some cloud along our line of sight in the hollow cone can explain the changing look of an AGN. Changes in energy generation are common, although what drives them is not understood.

\begin{figure}
\begin{center}
\floatbox[{\capbeside\thisfloatsetup{capbesideposition={left,top},capbesidewidth=4cm}}]{figure}[\FBwidth]
{\includegraphics[width=.3\textwidth]{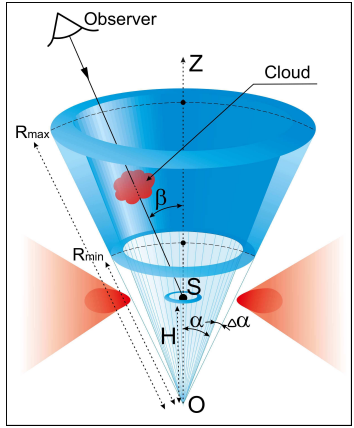}}
{\caption{Geometry of the near side of the proposed
hollow bi-conical dust distribution. Sublimation or recreation of dust in some cloud on the line of sight in the hollow can explain CL cases.}}
\label{fig1}
\end{center}
\end{figure}

\section{Monitoring of the typical CL object NGC 2617}
NGC 2617 is a typical example of a CL AGN. It has been observed to undergo a dramatic change from a largely obscured Seyfert 1.8 to an unobscured Seyfert 1 from 2003 to 2013 \cite{Shappee13, Shappee14}. We began spectroscopic and photometric (IR $JHK$ and optical $BVRI$) monitoring of NGC 2617 in January 2016 to see if it was still in a Seyfert 1 state (see details in \cite{Oknyansky17, Oknyansky17b, Oknyansky18}). In 2016 and 2017 NGC 2617 brightened again to a level of activity close to that in April 2013. However, from the beginning of April 2017 till the end of May 2018 it had a very low level of brightness and variability. In December 2017 the X-ray flux was the lowest since monitoring began in 1982. A similar deep minimum was observed in the optical in April--May 2018. Our most recent optical spectra obtained in April--May 2018 show a very low intensity of the broad H$\beta$ line.  Recently we reported \cite{Oknyansky18} a new brightening in the beginning of June 2018 after a very long low state during the preceding year (see details references above and related papers in preparation). Given the known low states in 2010, 2014 and 2017, the timescale of state changes is of the order of a few years. It would be interesting to know if this is quasi-periodic.  The possibility of quasi-periodical variability has been discussed for some other well-known CL AGNs such as NGC~4151 \cite{Oknyansky16d} and NGC~5548 \cite{Bon16}.
Such type cyclical periodicity is expected for TDEs \cite{Xiang16} or  tidal stars stripping \cite{Campana15, Ivanov-Chernyakova06}.\\

	NGC~2617 has very high amplitude X-ray variability (about a factor of 30).  This is similar to the X-ray variability of NLS1s \cite{Komossa17}, but a significant difference is that NLS1s mostly do {\em not} show so strong optical and UV variability (e.g. \cite{Klimek04}). As it has been shown \cite{Shappee14, Oknyansky17, Fausnaugh18} variability of NGC~2617 at all wavelengths is well correlated and the lags relative to X-ray variability increase with wavelength.  Typically, the UV lags the X-rays by about 1.5-3 days but optical variations lag the UV by only $\thickapprox 0.5$ days. The wavelength dependence of these lags is not consistent with the simplest ``lamp-post'' reprocessing models. Similar results were found in  another CL AGN, NGC~4151 \cite{Edelson17}.
For NLS1s the correlation between X-ray and UV variability is weak or non-existent for some objects (see e.g. \cite{Buisson18}).


\section{Changing-look NLS1s?}

Large-amplitude X-ray variability is a common property of NLS1s and CL AGNs.  The X-ray light curve of NGC~2617 (see Figure 4 of \cite{Shappee14} and \cite{Oknyansky18}) shows sudden drop outs and it has been proposed that drop outs in NLS1 X-ray variability could be due to occultation events \cite{Brandt+Gallagher00}. More work is needed to see how similar CL AGN and NLS1 light curves are.

To date there are only three reported cases of major optical changes in NLS1s: CSS100217 \cite{Drake11}, PS16dtm  \cite{Blanchard17}, and SDSS J123359.12+084211.5 \cite{Macleod17, Macleod18}. Probably these objects are examples of changing between a type-1 NLS1 and a
type-2 (i.e., hidden) NLS1.
The outbursts of these CL NLS1 are of {\em much} higher amplitude  than normal NLS1 variability.  The amplitudes of the outbursts, their durations, and shape of their decays \cite{Drake11,Blanchard17} all favor the outbursts arising from tidal disruption events (TDEs) instead (or possibly from supernovae).  Additional support for the outbursts being TDEs comes from the black-body shapes of the spectral energy distributions of the outburst components, and the absence of associated X-ray outbursts \cite{Blanchard17}.

TDEs and supernovae are rare.  However, it is not possible to rule them out on statistical grounds since there are only a few events  and they were found in targeted searches for TDEs and supernovae. However, if we take into account that historically NLS1s have not been monitored as well as BLS1s, CL variability masquerading as TDEs could be more common.
An important way of distinguishing  between different possible interpretations of CL NLS1 is trying to determine if these cases are just  one-off events.  As the case of NGC~2617 discussed above illustrates, recurrent brightening and fading is observed in CL BLS1s \cite{Oknyansky18, Komossa17}. It is therefore important to have long-term follow-up monitoring of objects like these CL NLS1 to see if outbursts repeat.

NGC~4051 is a typical NLS1 with strong X-ray variations. It has been reported to be a CL in X-rays \cite{uttley99} and it can be also identified as a CL AGN in the optical if we take into account the disappearance of broad He\,II $\lambda$4686 line at the time of a drop in X-ray flux \cite{Peterson00}. At the same time, the Balmer lines were several times narrower but did not disappear like He\,II $\lambda$4686. This difference can be explained by a simple model where Balmer lines arise primarily in a disk-like configuration seen at low inclination which is seen close to face-on in the case of NGC 4051, and the high-ionization lines (such as He\,II) arise primarily in an outflowing polar wind \cite{Gaskell82}.  In this picture the near-IR flux must be radiated from dust clouds which must also be located close to the plane of the accretion disk. In our model of hollow bi-conical dust outflow (see Fig.~1) the  free of dust region would have to have large half-opening angle, $\alpha$, for NGC 4051 and similar objects. When accretion rate onto the black hole is close to the Eddington limit (as is believed to be the case for NLS1s), the inner parts of the disc becomes thick \cite{shakura_sunyaev73}.  This can explain the flatter geometry of the BLR and dust region. Another reason for a flat dust geometry could be that inner disc truncation occurred because of a star transiting the disk \cite{Xiang16}. This special geometry of the dust region can be tested by IR reverberation mapping.

\section{Conclusions}
We must stress that, contrary to what is sometimes thought, the CL phenomenon is {\em not} rare among the BLS1s and QSOs.  A large fraction of strongly-variable AGNs will probably show type changes if observed enough \cite{Oknyansky17}. We note that there are similarities in X-ray variability of CL AGNs and NLS1s.  However, only a few cases of NLS1s are known to change  of their optical spectra. These objects (as well as all NLS1s) need much more intensive monitoring  to understand better the physics of such changes.

As mentioned, CL AGNs present problems for the simplest unification models. Orientation obviously cannot change on the time-scale of the observed type changes, and hence some other explanation is needed.  CL AGNs cases can be explained with changing obscuration and changing of accretion rates.  We suspected that the underlying reason is strong variations of the accretion rate which in turn might change the obscuration.  What must happen to make such a dramatic changes possible?  The main problem with invoking a TDE or supernova near SMBH in some AGN is that they are extremely rare. Repeat tidal stars stripping \cite{Campana15, Ivanov-Chernyakova06} has higher  probability \cite{Komossa17}, but this possibility is not investigated enough yet.

\section*{Acknowledgements}

It is our pleasure to thank the organizers of the conference for their invitation and for their excellent organization and the pleasant atmosphere. We are most grateful to Dr. Chelsea MacLeod for sending us unpublished information about SDSS J123359.12+084211.5. We thank graphic designer Natalia Sinugina for help with Fig.1. VLO acknowledges financial support from the RadioNet which has received funding from the European Union's Horizon 2020 research and innovation programme under grant agreement No~730562. KLM acknowledges support from RBFR grant 18-32-00553.



\bibliography{mybib}

\end{document}